\documentstyle[twocolumn,psbox,myletter]{jpsj}

\def\runtitle{Two-Triplet-Dimer Excitation Spectra in the Shastry-Sutherland Model}

\def\runauthor{Yoshiyuki {\sc Fukumoto}}

\title{Two-Triplet-Dimer Excitation Spectra in 
the Shastry-Sutherland Model for ${\bf SrCu_2(BO_3)_2}$}

\author{Yoshiyuki {\sc Fukumoto}\footnote{E-mail: yfuku@ph.noda.sut.ac.jp}}

\inst{Department of Physics, Faculty of Science and Technology, Science University of Tokyo,\\
Noda, Chiba 278-8510\\}

\recdate{\hspace*{3cm}}

\abst{By using the perturbation expansion up to the fifth order, we study the
two-triplet-dimer excitation spectra in the Shastry-Sutherland model, where 
the localized nature of a triplet-dimer, the propagation of a triplet-dimer pair 
by the correlated hopping and the long-range interactions between triplet-dimers 
play an essential role. It is found that the dispersion relations for 
first-neighbor triplet-dimer pair excitations with $S=1$ and $p$-type symmetry 
qualitatively explain the second-lowest branch observed in the neutron inelastic scattering 
experiment. It is also predicted that the second-lowest branch consists of two components, 
$p_x$- and $p_y$-states, with slightly different excitation energies. The origin of the
singlet mode at 3.7meV observed in the Raman scattering experiment is also discussed.}

\kword{${\sf SrCu_2(BO_3)_2}$, Shastry-Sutherland model, two dimensions, bound state, 
Heisenberg antiferromagnet, perturbation expansion, dispersion relation}

\begin{document}
\sloppy
\maketitle

There has been a growing interest in low-dimensional quantum spin systems, 
since one can observe a variety of properties where classical pictures break down.
In the last decade, spin gapped ground states, particularly, in two-dimensional systems
have received considerable attention in connection with the high-$T_{\rm c}$ 
superconductivity. For instance, a quasi-two-dimensional compound ${\rm CaV_4O_9}$ 
is known to have a spin gap originating in the plaquette RVB 
mechanism.~\cite{rf:Taniguchi,rf:Kodama,rf:Ueda,rf:Fukumoto1}

\begin{figure}
\begin{center}
\psbox[scale=1.0#1]{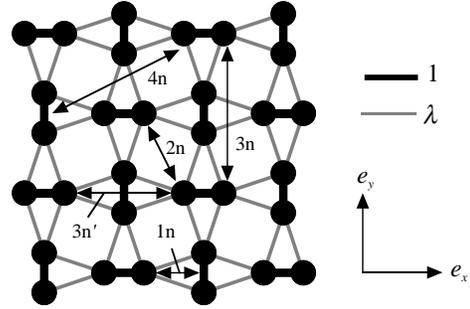}
\end{center}
\caption{Structure of a two-dimensional network formed by ${\rm Cu}^{2+}$ in 
${\rm SrCu_2(BO_3)_2}$. The closed circles represent copper ions. The nearest-neighbor
bonds are expressed by the bold lines and the next nearest-neighbor bonds by 
the gray lines. The vectors, ${\mib e}_x$ and ${\mib e}_y$, are the primitive translation 
vectors in this lattice. Note that there exist two types of third-neighbor triplet-dimer pairs,
which are indicated by 3n and ${\rm 3n}^\prime$. For brevity, we call the farmer the
third-neighbor pair. Within the fifth-order calculation, the diagonal
interactions between triplet-dimers appear for the first-neighbor (1n),  
second-neighbor (2n), third-neighbor (3n) and fourth-neighbor (4n) triplet-dimer pairs.} 
\label{fig:structure}
\end{figure}

Two years ago, another new two-dimensional spin gap system ${\rm SrCu_2(BO_3)_2}$
was found by Kageyama {\it et al.}~\cite{rf:Kageyama1} In this compound, magnetic 
ions ${\rm Cu}^{2+}$ ($S=1/2$) are arranged as shown in Fig.~\ref{fig:structure}.
Miyahara and Ueda~\cite{rf:Miyahara1} pointed out that ${\rm SrCu_2(BO_3)_2}$ is an
experimental realization of a special class of the Heisenberg antiferromagnets
which is called the Shastry-Sutherland model~\cite{rf:Shastry}
\begin{equation}
\label{eq:H}
   {\cal H} = \sum_{\left<i,j\right>}{\mib S}_{i}\cdot {\mib S}_{j}+
              \lambda\sum_{\left<\left<i,j\right>\right>}{\mib S}_{i}\cdot {\mib S}_{j},
\end{equation}
where $\left<i,j\right>$ ($\left<\left<i,j\right>\right>$) denotes a (next-)nearest-neighbor 
pair of spins. The strength of the interdimer coupling is now considered 
$\lambda\sim 0.63$.~\cite{rf:Miyahara3,rf:Weihong}
The orthogonal dimer structure as seen in Fig.~\ref{fig:structure} leads to the following
unique properties: the direct product of the singlet-dimer states of the nearest-neighbor 
pairs of spins is the ground state exactly for $\lambda<\sim 0.7$ and a triplet-dimer 
in the singlet sea is almost localized.~\cite{rf:Miyahara1,rf:Weihong} 

The nature of the quantum phase transition has been one of important subjects in theoretical 
investigations.~\cite{rf:Miyahara1,rf:Weihong,rf:Albrecht,rf:Muller,rf:Koga}
At first stage of the investigations, it was assumed that the exact dimer singlet state 
is destabilized against the N\'{e}el ordered state when $\lambda$ is increased.
A recent study, however, suggested that there exists an intermediate phase, where
a plaquette RVB state is stabilized, between the exact dimer singlet phase and the N\'{e}el 
ordered phase.~\cite{rf:Koga} The first-order transition point between the 
exact dimer singlet phase and the plaquette RVB phase was estimated as $\lambda_{\rm c}=0.677$
by using the series expansion method. Another interesting topics in ${\rm SrCu_2(BO_3)_2}$ 
is that the magnetization plateaus are observed at 1/3, 1/4 and 1/8 of the full Cu 
moment.~\cite{rf:Kageyama1,rf:Kageyama3} The localized nature of a triplet-dimer 
plays an important role in the appearance of the 
plateaus.~\cite{rf:Miyahara1,rf:Momoi,rf:Miyahara2,rf:Fukumoto2}

As mentioned above, a triplet-dimer is hard to propagate in the singlet sea because of 
the orthogonal dimer structure. In precise, the orthogonal dimer structure prohibits 
the propagation of a triplet-dimer up to the fifth-order perturbation in 
$\lambda$.~\cite{rf:Miyahara1,rf:Miyahara2} Recently, almost flat band of the 
single-triplet-dimer excitations at 3.0meV was observed in the neutron inelastic 
scattering experiment by Kageyama {\it et al.}~\cite{rf:Kageyama2}

In this neutron inelastic scattering experiment, multiple-triplet-dimer excitation 
spectra were also observed, which show rather dispersive behavior compared with the 
single-triplet-dimer excitation. In the ESR measurement, the second spin-gap was observed 
at 4.7meV, which corresponds to the band bottom of the second-lowest spectrum.~\cite{rf:Nojiri}
Note that the localized nature of a triplet-dimer leads to almost flat bands of
multiple-triplet-dimer continua and makes the experimental observation of 
various multiple-triplet-dimer bound states possible. The observation of a singlet bound 
state at 3.7meV in the Raman scattering experiment has been also reported.~\cite{rf:Lemmens}

In this letter, we study the two-triplet-dimer excitations by the fifth-order perturbation 
expansion. We here mention the relation between the present problem and the two-magnon bound 
state in the Ising-Heisenberg ferromagnets which is now well understood.~\cite{rf:book}
In the Ising-Heisenberg model, when the hopping of a magnon is suppressed by the Ising-like 
exchange anisotropy, the formation of the two-magnon bound states is more favored and 
the band of those states becomes more narrow. Within the first-order perturbation, the 
mechanism of the formation of two-triplet-dimer bound states in the present model is the 
same as that of the the two-magnon bound states in the Ising limit of 
the Ising-Heisenberg model.

In the Shastry-Sutherland model, the two-triplet-dimer bound state problem becomes to 
be more interesting due to the following two reasons arising from higher-order 
perturbation processes. First, the range of the diagonal interactions between 
triplet-dimers is enlarged when the order of the perturbation is extended. Second,
the correlated hopping, where a triplet dimer can hop when the rest spectator 
triplet-dimer always lies at a neighboring site before and after the hopping, 
is possible in the lower-order perturbation processes than the six order which leads to the
single-triplet-dimer hopping.~\cite{rf:Momoi,rf:Miyahara2,rf:Fukumoto2}
The former separates a branch of two-triplet-dimer bound states from the two-triplet-dimer 
continuum. The latter make such the branch a dispersive one. It is also important 
to observe these higher-order effects pronouncedly that ${\rm SrCu_2(BO_3)_2}$ locates 
near the first-order transition point. We expect that the two facts mentioned above and the 
localized nature of a triplet-dimer are properly taken into account in our fifth-order 
perturbation calculation.

We now turn to the calculation of matrix elements in the fifth-order effective Hamiltonian, 
$H^{\rm eff}$, for the subspace composed of two triplet-dimers. We write the two-triplet-dimer 
states as
\begin{equation}
\label{eq:base1}
   \left|\mib{r},\mib{\delta},S \right>\equiv \sum_{m=-1}^{1}C(S,m)
   \left|t_m \right>_{\mibs r}\left|t_{-m} \right>_{{\mibs r}+\mibs{\delta}}
   {\prod_{{\mibs r}^{\prime}}}^{\prime}\left|s \right>_{{\mibs r}^{\prime}},
\end{equation}
where $S=0,1,2$. In this definition, $\left|s \right>_{\mibs r}$ represents the singlet 
state of the dimer at ${\mib r}$, and $\left|t_m \right>_{\mibs r}$ represents the 
triplet state with the total $S^z$ of $m$. The prime attached on the product 
symbol means the exception of the sites ${\mib r}$ and ${\mib r}+\mib{\delta}$. 
The explicit expression of $C(S,m)$ is given by
\begin{equation}
   C=\left[ \begin{array}{ccc}
               -1/\sqrt{3} & 1/\sqrt{3} & -1/\sqrt{3}\\
                1/\sqrt{2} &          0 & -1/\sqrt{2}\\
                1/\sqrt{6} & 2/\sqrt{6} &  1/\sqrt{6}
               \end{array} \right ],
\end{equation}
where the row (column) index runs $S=0,1,2$ ($m=-1,0,1$). Note that there is a symmetry
relation $C(S,-m)=(-1)^S C(S,m)$. The relative vector between the two triplet-dimers, 
${\mib \delta}$, is assumed to satisfy $\delta_x>0$ or $\delta_x=0,\;\delta_y>0$ in 
the practical calculations. 

The interactions between two triplet-dimers within fifth-order approximation are given by
\begin{equation}
   V_{\mibs{r},\mibs{\delta}}(S)=
   \left<\mib{r},\mib{\delta},S\right|H^{\rm eff}\left|\mib{r},
   \mib{\delta},S\right>-\left(E_{\rm g}+2 \Delta_{\rm sg}\right),
\end{equation}
where $E_{\rm g}=-3N_{\rm D}/4$ is the ground state energy of the system with
$N_{\rm D}$ dimers and
\begin{equation}
   \Delta_{\rm sg}=1-\lambda^2-\frac{\lambda^3}{2}-\frac{\lambda^4}{8}+
                             \frac{5\lambda^5}{32}
\end{equation}
is the fifth-order series of the spin gap.~\cite{rf:Miyahara1,rf:Weihong} Nonzero elements of 
$V_{\mibs{r},\mibs{\delta}}(S)$ are summarized in Table~\ref{table:values}. The
interactions for $S=2$ have been calculated already up to the third order and the 
first-order repulsion between the first-neighbor triplet-dimers is known to lead to the 
1/2-plateau.~\cite{rf:Momoi,rf:Miyahara2,rf:Fukumoto2} On the other hand, we find in
Table~\ref{table:values} that the first-order interactions between first-neighbor 
triplet-dimers for $S=0$ and 1 are attractive.

\begin{table}[h]
\centering
\caption{Nonzero elements of the diagonal interactions between two triplet-dimers,
$V_{\mibs{r},\mibs{\delta}}(S)$. \hspace{-1mm}A set of (\mib{r},\mib{\delta}) for first-,
second-, third- or fourth-neighbor triplet-dimer pairs is, respectively, 
denoted by 1n, 2n, 3n or 4n. (See Fig.~\ref{fig:structure}.)}
\label{table:values}
\begin{tabular}{@{\hspace{\tabcolsep}\extracolsep{\fill}}ccc}
\hline
$(\mib{r},\mib{\delta})$ & $S$ & $V_{\mibs{r},\mibs{\delta}}(S)$ \\
\hline
   & 0 & $\lambda(-16+8\lambda+16\lambda^2+18\lambda^3-\lambda^4)/16$  \\
1n & 1 & $\lambda(-8+16\lambda+14\lambda^2-9\lambda^3-16\lambda^4)/16$ \\
   & 2 & $\lambda(8+8\lambda-2\lambda^2-9\lambda^3-\lambda^4)/16$      \\
\hline
   & 0 & $\lambda^3(-16+45\lambda^2)/32$        \\
2n & 1 & $\lambda^3(-4-6\lambda+3\lambda^2)/16$ \\
   & 2 & $\lambda^3(2+3\lambda+3\lambda^2)/8$   \\
\hline
   & 0 & $\lambda^2(-32-48\lambda+56\lambda^2+289\lambda^3)/32$ \\
3n & 1 & $\lambda^2(-32-48\lambda+8\lambda^2+129\lambda^3)/64$  \\
   & 2 & $\lambda^2(32+48\lambda+8\lambda^2+129\lambda^3)/64$   \\
\hline
   & 0 & $\lambda^4(-8-17\lambda)/32$ \\
4n & 1 & $\lambda^4(-8-17\lambda)/64$ \\
   & 2 & $\lambda^4(8+17\lambda)/64$  \\
\hline
\end{tabular}
\end{table}

We introduce the Fourier transformation of eq.~(\ref{eq:base1}):
\begin{equation}
   \left|\mib{q},\mib{\delta},S,\alpha\right>=\sqrt{\frac{2}{N_{\rm D}}}
   \sum_{\mibs{r}_{\alpha}}{\rm e}^{{\rm i}\mibs{q}\cdot\mibs{r}_\alpha}
   \left|\mib{r}_\alpha,\mib{\delta},S\right>,
\end{equation}
where $\alpha(=A,B)$ denotes the sublattice index and $\mib{r}_\alpha$ denotes a site on 
the sublattice $\alpha$. The representation of $H^{\rm eff}$ using this basis set becomes 
to be a direct sum of finite-size matrices because of the prohibition of the 
single-triplet-dimer hopping. In the fifth-order approximation, the representation of 
$H^{\rm eff}$ has off-diagonal matrix elements among first-, second-, fourth-neighbor 
triplet-dimer pair states. For the third-neighbor triplet-dimer pairs, 
$H^{\rm eff}$ has only a diagonal element, which is different from 
$E_{\rm g}+2 \Delta_{\rm sg}$. For the other states, $H^{\rm eff}$ has only the diagonal 
element of $E_{\rm g}+2 \Delta_{\rm sg}$.

We here mention the zeroth-order wave-functions for the first-neighbor 
triplet-dimer pair excitations, which are relevant to the observed multiple-triplet-dimer
excitations in the neutron inelastic scattering experiment
as described below. The first-order perturbation removes the degeneracy between the 
first-neighbor triplet-dimer pair states and the other states. We write the eigenfunctions 
of the eigenvalue problem to determine the second-order energies as follows:
\begin{subequations}
\begin{equation}
   \left|s_{\xi}\right>_{S,\mibs{q}}\equiv
   \frac{{\rm e}^{-{\rm i}q_{\xi}}}{\sqrt{2}}\left|{\mib q},\mib{e}_{\xi},S,\alpha_{\xi}\right>
   +\frac{(-1)^S}{\sqrt{2}} \left|{\mib q},\mib{e}_{\xi},S,\bar{\alpha}_{\xi}\right>,
\end{equation}   
\begin{equation}
   \left|p_{\xi}\right>_{S,\mibs{q}}\equiv
   \frac{{\rm e}^{-{\rm i}q_{\xi}}}{\sqrt{2}}\left|{\mib q},\mib{e}_{\xi},S,\alpha_{\xi}\right>
   -\frac{(-1)^S}{\sqrt{2}} \left|{\mib q},\mib{e}_{\xi},S,\bar{\alpha}_{\xi}\right>
\end{equation}   
\end{subequations}
for $\xi=x,y$, where we have defined $\alpha_{x}\equiv A$, $\alpha_{y}\equiv B$, 
$\bar{\alpha}_{x}\equiv B$ and $\bar{\alpha}_{y}\equiv A$. 
Note that $\left|s_{\xi}\right>_{S,0}$ 
and $\left|p_{\xi}\right>_{S,0}$ have, respectively, the $s$-type and $p_{\xi}$-type 
symmetries. The degeneracy between the $s$-wave states and the $p$-wave states is removed by 
the second-order perturbation. The fourth-order perturbation removes the remaining two-fold 
degeneracy for each of the $s$-wave states and the $p$-wave states at the general points 
in the first Brillouin zone. Then the way of mixing between the two $s$-wave functions 
is determined as follows:
\begin{subequations}
\begin{equation}
   \left|s_{+}\right>_{S,\mibs{q}}\equiv
      f(q_x,q_y)\left|s_x\right>_{S,\mibs{q}}+
      (-1)^S f(q_y,q_x)\left|s_y\right>_{S,\mibs{q}},
\end{equation}   
\begin{equation}
   \left|s_{-}\right>_{S,\mibs{q}}\equiv
      f(q_y,q_x)\left|s_x\right>_{S,\mibs{q}}-
      (-1)^S f(q_x,q_y)\left|s_y\right>_{S,\mibs{q}},
\end{equation}   
\end{subequations}
where
\begin{eqnarray}
   &\left[f(q_x,q_y)\right]^{-2}&=1+ \nonumber \\
   &&\hspace*{-1.3cm}\left[\cos q_x-\cos q_y-\sqrt{1+(\cos q_x-\cos q_y)^2}\right]^2.
\end{eqnarray}

\begin{figure}
\begin{center}
\psbox[scale=1.0#1]{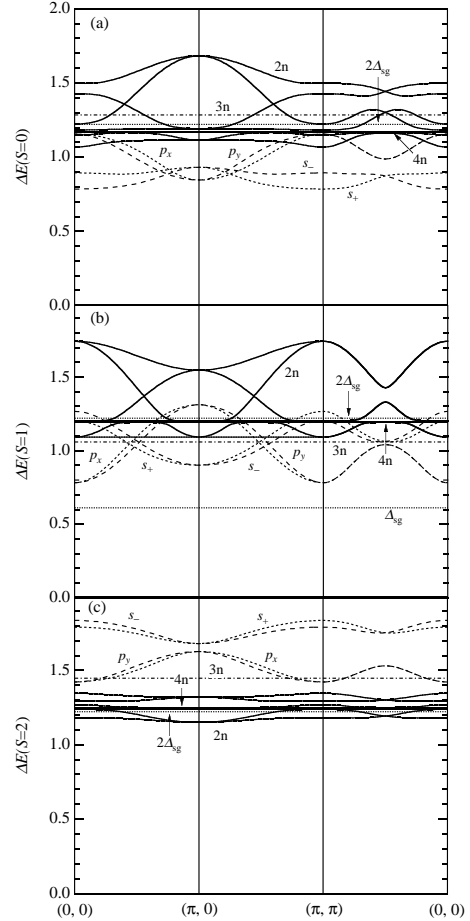}
\end{center}
\caption{Dispersion relations of the two-triplet-dimer states 
for (a) $S=0$, (b) $S=1$ and (c) $S=2$, where we choose 
$\lambda=0.55$. For $S=1$, dispersion relation of the single-triplet-dimer 
states is also shown.} 
\label{fig:dispersion}
\end{figure}

The results of the fifth-order perturbation expansion of excitation energies, $\Delta E$,
for all the two-triplet-dimer excitations are shown in Fig.~\ref{fig:dispersion},
where we choose $\lambda=0.55$. In this figure, the results except for the second- and 
fourth-neighbor triplet-dimer pairs are obtained by the usual series expansion.
For the second- and fourth-neighbor triplet-dimer pairs, the mixing among these states 
at high-symmetry points of ${\mib q}$ occurs in the fifth order, which makes 
the convergence of the series slow around such the high-symmetry points. To avoid
this, we diagonalize directly the effective Hamiltonian for the subspace of the 
second- and fourth-neighbor triplet-dimer pair states, which is an hermitian matrix obtained 
by making further partial diagonalization for $H^{\rm eff}$. The resultant eigenvalues are 
plotted in Fig.~\ref{fig:dispersion} as the solid lines. Roughly speaking, almost 
dispersionless bands indicated by "4n" originate in the fourth-neighbor triplet-dimer pairs. 
The other solid lines originate in the second-neighbor triplet-dimer pairs.

We find dispersive behavior for the branches of the first-neighbor triplet-dimer 
pairs for $S=0$, 1 and 2, which can propagate in the singlet sea from the third order.
It should be stressed that the dispersion relation of the first-neighbor triplet-dimer pairs
with $p$-type symmetry for $S=1$ qualitatively explains the behavior of the second-lowest
branch in the neutron scattering experiment. (See Fig.~3 in Ref.~17.)
Very recently, a neutron inelastic scattering experiment under the magnetic fields was carried
out.~\cite{rf:Kakurai} This experiment revealed that the total spin of the second-lowest 
branch is one, which is also consistent with our result. Note that the fourth-order terms 
remove the two-fold degeneracy between $p_x$- and $p_y$-type first-neighbor two-triplet-dimer 
pairs at general points in the first Brillouin zone.
The explicit expressions of the excitation energies are given by
\begin{subequations}
\begin{eqnarray}
\label{eq:px}
   \Delta E_{p_x}(S=1)&=&2-\frac{\lambda}{2}-\frac{3\lambda^2}{4}
   -\frac{\lambda^3}{4}(1+\cos q_x\cos q_y)\nonumber \\
   &&\hspace*{-13mm}-\frac{\lambda^4}{16}
     (35+15\cos q_x \cos q_y-2\cos^2 q_y \sin^2 q_x) \nonumber \\
   &&\hspace*{-13mm}-\frac{\lambda^5}{64}\left[274+
     175\cos q_x \cos q_y-8\cos^2 q_x \right.\nonumber \\
   &&\hspace*{-5mm}\left.\times \left\{1+(6+\cos q_x \cos q_y)\sin^2 q_y\right \}\right],
\end{eqnarray}
\begin{equation}
   \Delta E_{p_y}(S=1)=(q_x\leftrightarrow q_y\;\mbox{in rhs of eq.~(\ref{eq:px})}),
\end{equation}
\end{subequations}
for the $p_x$- and $p_y$-states, respectively.
As seen in the above equations and Fig.~\ref{fig:dispersion}(b), the observed 
second-lowest branch is composed of two components with slightly different energies
on the path from the $\Gamma$ point to the X point and on the path from the X point 
to the M point, which may be detected by careful analysis of the 
experimental data.

We turn to the second-neighbor triplet-dimer pairs. From Fig.~\ref{fig:dispersion}, 
we find dispersive behavior for $S=0$ and 1, but we find rather flat bands for $S=2$. 
Note that the effective Hamiltonian for $S=2$ is nothing but the two-body part in the 
subspace of the effective Hamiltonian dominating the magnetization process. Rather small 
energy-gain by the correlated hopping for $S=2$ seems to be consistent 
with the appearance of the magnetization plateaus in ${\rm SrCu_2(BO_3)_2}$. 
In Ref.~16, where the magnetization plateaus are studied by the third-order perturbation,
it was assumed that the states with first-neighbor triplet-dimer pairs can be 
truncated. We find in Fig.~\ref{fig:dispersion}(c) that the excitation energies of the 
first-neighbor two-triplet-dimer pairs for $S=2$ are not the lowest lying branch
even when the fourth- and fifth-order terms are taken into account.
This finding gives a support to the treatment in Ref.~16.

As for the third-neighbor triplet-dimer pairs, the binding energies are not so 
small, but propagation does not occur within the fifth-order perturbation. 
The fourth-neighbor triplet-dimer pairs can propagate from the fifth order but
the band width is too narrow to see in this figure.

\begin{figure}
\begin{center}
\psbox[scale=1.0#1]{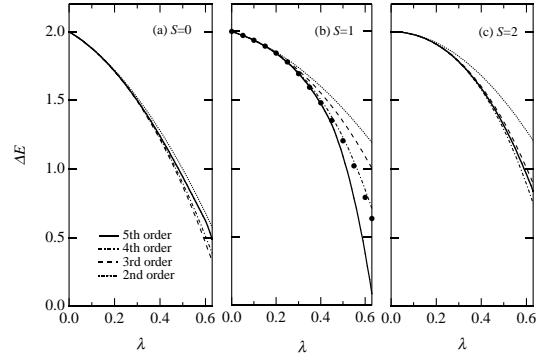}
\end{center}
\caption{The minimum two-triplet-dimer excitation energies as a function of $\lambda$
for (a) $S=0$, (b) $S=1$ and (c) $S=2$. The solid, dash-dotted, long-dashed and 
short-dashed lines, respectively, represent the fifth-, fourth-, third- and second-order 
results. The closed circles in (b) are the results of the exact diagonalization for 
the finite-size cluster with $N_{\rm D}=10$.} 
\label{fig:ed}
\end{figure}

In Fig.~\ref{fig:ed}, we show the minimum two-triplet-dimer excitation energies
as a function of $\lambda$. For $S=1$, the result of the exact diagonalization
on a finite-size cluster with $N_{\rm D}=10$ is also shown. We find that the series of the 
two-triplet-dimer excitation for $S=0$ and $S=2$ tends to converge within the present 
calculation. In particular, the fifth-order perturbation series results in 3.7meV for
the minimum energy of the singlet two-triplet-dimer excitation for 
${\rm SrCu_2(BO_3)_2}$ with $\lambda=0.63$ and the intradimer exchange constant of 7.3meV.
This value agrees with the observed singlet excitation energy in the Raman scattering 
experiment.~\cite{rf:Lemmens} As for $S=1$, the result in Fig.~\ref{fig:ed}(b) 
indicates that we should extend the order of perturbation to obtain the quantitative
description of ${\rm SrCu_2(BO_3)_2}$. Note that the exact diagonalization 
gives 4.6meV for the minimum two-triplet-dimer excitation energy for ${\rm SrCu_2(BO_3)_2}$,
which agrees with the results of the neutron inelastic scattering 
experiment~\cite{rf:Kageyama2} and the ESR measurement.~\cite{rf:Nojiri}
Our fifth-order result of this excitation energy is lower than that of the exact 
diagonalization. If higher-order terms beyond the fifth order are taken into account, 
then the binding energy is expected to be suppressed due to the single-triplet-dimer hopping.

In summary, we have studied the two-triplet-dimer excitations in the Shastry-Sutherland 
model by using the fifth-order perturbation. It has been found that the second-lowest branch 
observed in the neutron scattering experiment and the singlet state at 3.7meV observed 
in the Raman scattering experiment originate from the first-neighbor two-triplet-dimer 
excitations. It has been also pointed out that the second-lowest branch in the neutron 
scattering experiment is composed of two components, $p_x$- and $p_y$-states, whose degeneracy
is removed by the fourth-order perturbation. We expect that such the structure is confirmed by
the careful analysis of the experimental data.

The author would like to thank Prof. A. Oguchi for useful discussions and critical 
reading of the manuscript. We have used a part of the codes provided by H. Nishimori in 
TITPACK Ver.~2.


\begin{thebibliography}{99}
\bibitem{rf:Taniguchi} S. Taniguchi, T. Nishikawa, Y. Yasui, Y. Kobayashi, 
M. Sato, T. Nishioka, M. Kontani and K. Sano: J. Phys. Soc. Jpn. {\bf 64} (1995) 2758.
\bibitem{rf:Kodama} K. Kodama, H. Harashina, H. Sasaki, Y. Kobayashi, M. Kasai, 
S. Taniguchi, Y. Yasui, M. Sato, K. Kakurai, T. Mori and M. Nishi: 
J. Phys. Soc. Jpn. {\bf 66} (1997) 793.
\bibitem{rf:Ueda} K. Ueda, H. Kontani, M. Sigrist and P. A. Lee: Phys. Rev. Lett. 
{\bf 76} (1996) 1932.
\bibitem{rf:Fukumoto1} Y. Fukumoto and A. Oguchi: J. Phys. Soc. Jpn. {\bf 67} (1998) 2205.
\bibitem{rf:Kageyama1} H. Kageyama, K. Yoshimura, R. Stern, N. V. Mushnikov, K. Onizuka, 
M. Kato, K. Kosuge, C. P. Slichter, T. Goto and Y. Ueda: Phys. Rev. Lett. {\bf 82} (1999) 3168.
\bibitem{rf:Miyahara1} S. Miyahara and K. Ueda: Phys. Rev. Lett. {\bf 82} (1999) 3701.
\bibitem{rf:Shastry} S. Shastry and B. Sutherland: Physica {\bf 108}B (1981) 1069.
\bibitem{rf:Miyahara3} S. Miyahara and K. Ueda: preprint.
\bibitem{rf:Weihong} Z. Weihong, C. J. Hamer and J. Oitmaa: Phys. Rev. B{\bf 60} (1999) 6608.
\bibitem{rf:Albrecht} M. Albrecht and F. Mila: Europhys. Lett. {\bf 34} (1996) 145.
\bibitem{rf:Muller} E. Muller-Hartmann, R. R. P. Singh, C. Knetter and G. S. Uhrig: 
cond-mat/9910165.
\bibitem{rf:Koga} A. Koga and N. Kawakami: preprint.
\bibitem{rf:Kageyama3} H. Kageyama, Y. Narumi, K. Kindo, K. Onizuka, Y. Ueda and T. Goto: 
preprint.
\bibitem{rf:Momoi} T. Momoi and K. Totsuka: Phys. Rev. B{\bf 61} (2000) 3231.
\bibitem{rf:Miyahara2} S. Miyahara and K. Ueda: Phys. Rev. B{\bf 61} (2000) 3417.
\bibitem{rf:Fukumoto2} Y. Fukumoto and A. Oguchi: J. Phys. Soc. Jpn. (in press.)
\bibitem{rf:Kageyama2} H. Kageyama, M. Nishi, N. Aso, K. Onizuka, 
T. Yoshimura, K. Nukui, K. Kodama, K. Kakurai and Y. Ueda: preprint.
\bibitem{rf:Nojiri} H. Nojiri, H. Kageyama, K. Onizuka, Y. Ueda and M. Motokawa: 
J. Phys. Soc. Jpn. {\bf 68} (1999) 2906.
\bibitem{rf:Lemmens} P. Lemmens, M. Grove, M. Fischer, G. Guntherodt, V. N. Kotov,
H. Kageyama, K. Onizuka and Y. Ueda: cond-mat/0003094.
\bibitem{rf:book} I. Ono, S. Mikado and T. Oguchi: J. Phys. Soc. Jpn. {\bf 30} (1971) 358.
\bibitem{rf:Kakurai} K. Kakurai: private communication.
\end{thebibliography}
\end{document}